\documentclass{article}
\usepackage{spconf,amsmath,epsfig}
\usepackage{hyperref}
\usepackage{graphicx}
\usepackage{algorithmic}
\usepackage{textcomp}
\usepackage{xcolor}
\usepackage{booktabs}

\let\OLDthebibliography\thebibliography
\renewcommand\thebibliography[1]{
  \OLDthebibliography{#1}
  \setlength{\parskip}{0pt}
  \setlength{\itemsep}{0pt plus 0.3ex}
}

\begin{document}\sloppy
\ninept

\def\x{{\mathbf x}}
\def\L{{\cal L}}

\title{Boosting Multi-Speaker Expressive Speech Synthesis with Semi-supervised Contrastive Learning}
%

\name{Xinfa Zhu$^1$, Yuke Li$^1$, Yi Lei$^1$, Ning Jiang$^2$, Guoqing Zhao$^2$, Lei Xie$^1$$^{*}$\thanks{* Corresponding author.}}
\address{
  $^1$Audio, Speech and Language Processing Group (ASLP@NPU)\\School of Computer Science,
  Northwestern Polytechnical University, Xi’an, China \\
  $^2$Mashang Consumer Finance Co., Ltd.}

\maketitle

\begin{abstract}
This paper aims to build a multi-speaker expressive TTS system, synthesizing a target speaker's speech with multiple styles and emotions.
To this end, we propose a novel \textit{contrastive learning}-based TTS approach to transfer style and emotion across speakers. Specifically, contrastive learning from different levels, i.e. \textit{utterance} and \textit{category} level, is leveraged to extract the disentangled style, emotion, and speaker representations from speech for style and emotion transfer.
Furthermore, a semi-supervised training strategy is introduced to improve the data utilization efficiency by involving multi-domain data, including style-labeled data, emotion-labeled data, and abundant unlabeled data. To achieve expressive speech with diverse styles and emotions for a target speaker, the learned disentangled representations are integrated into an improved VITS model. Experiments on multi-domain data demonstrate the effectiveness of the proposed method.  

\end{abstract}
\section{Introduction}
\label{sec:intro}

In recent years, neural text-to-speech (TTS) synthesis has rapidly progressed in speech quality and naturalness~\cite{Tan2021ASO,Ren2021FastSpeech2F,Kim2021ConditionalVA}. With the wide applications of TTS, there have been increasing demands for expressive speech synthesis systems to provide more human-like speech in diverse scenarios. Transfer learning has been the favored method for expressive speech synthesis, which aims to transfer expressiveness from speech recorded by other speakers to the target speaker~\cite{Sorin2020, Lei2022MsEmoTTSME}.

In previous works of expressive speech synthesis, speech expressiveness usually refers to specific speaking styles or emotional expressions. These works usually leverage a reference encoder to extract style or emotional expression from reference speech~\cite{Min2021MetaStyleSpeechM,DBLP:conf/iscslp/LiYXX21}. The critical factor for style or emotion transfer is to decouple the speaker timbre and expressive aspects from speech, as their entanglement usually leads to low speaker timbre similarity or expressiveness~\cite{Li2022CrossSpeakerED} for the generated speech. There are many approaches to achieve disentanglement, such as domain adversarial training (DAT)~\cite{DBLP:journals/jmlr/GaninUAGLLML16}, mutual information (MI)~\cite{Wang2021VQMIVCVQ}, and information perturbation~\cite{9874835}. Although researchers utilizing the above approaches have achieved good performance on style or emotion transfer, they have unclear definitions of style and emotion and sometimes consider emotion as a type of speaking style. This indiscriminate treatment of style and emotion restricts them from being extended to real scenarios requiring the combination of different emotions and styles. 

Therefore, this paper considers transfer style and emotion from different reference speech simultaneously following the prior investigation~\cite{10095776}, which has pointed out that \textit{speaking style} is a general distinctive style of speech in different usage scenarios, such as news reading, storytelling, poetry recitation, and spontaneous conversation. By contrast, \textit{emotion} mainly reflects the mood state of the speaker, related to attitudes and intentions, such as happy, angry, and sad. Building a system that focuses on both speaking style and emotion transfer in multi-speaker expressive speech synthesis usually requires multi-domain datasets containing diverse styles, emotions, and speakers. Meanwhile, speaking style, emotion, and speaker timbre are highly entangled as they all affect the prosody patterns of speech. Even in a reference-based model, the linguistic content of reference speech and texts is consistent during training, which causes entanglement of linguistic content and performance degradation in inference~\cite{DBLP:conf/interspeech/MengL0LSXSZM22}, further increasing the difficulty of disentanglement. To solve these problems, this previous study~\cite{10095776} designed a two-stage framework to conduct the disentanglement with neural bottleneck~(BN) features as the intermediate representation. Specifically, it proposed a sophisticated disentanglement mechanism by extracting a style/emotion/speaker representation for each category. The disentangled style/emotion/speaker attribute is treated as temporal irrelevant since it is aggregated to a fixed-length vector for the same category. With this assumption, different segments of the same utterance (utterance level) could be involved in extracting the disentangled representations, but this has not been explored in previous works. Moreover, this two-stage framework has many subsystems in which the error accumulation of BN prediction results in the degradation of synthetic speech naturalness.

Contrastive learning is a method to learn the desired features of the data via constructing positive and negative samples. Recently, approaches based on contrastive learning have shown remarkable performance in many fields, such as computer vision (SIMCLR~\cite{SIMCLR}, CLIP~\cite{CLIP}), reinforcement learning (CURL~\cite{CULR}), and speech processing (MULAN~\cite{Mulan}, CLAPSpeech~\cite{CLAPSpeech}). Besides, contrastive learning has shown superiority in speech emotion recognition~\cite{scser,uscser}, achieving state-of-the-art performance. By respectively minimizing and maximizing the distances of positive and negative samples~\cite{surveyoncl}, contrastive learning has the advantage of extracting latent representations for specific attributes. Besides, through constructing sample pairs from multiple levels, contrastive learning exhibits the potential for performance improvements~\cite{multilevels}.

With the above considerations, this paper investigates the effectiveness of contrastive learning in the challenging expressive TTS task that enables multi-speaker, multi-style, and multi-emotion speech synthesis. Specifically, this paper proposes a novel contrastive learning-based TTS approach to transfer style and emotion across speakers by extracting the desired attribute representations (i.e., style, emotion, and speaker in this paper). First, we design a Speech Representation Learning (SRL) module to extract style/emotion/speaker-only-related vectors by conducting contrastive learning from both utterance level and category level.
Second, we introduce a semi-supervised training
strategy to the SRL module, which can effectively leverage multi-domain speech data, including style-labeled, emotion-labeled, and unlabeled data. With this strategy, the SRL can be trained on abundant speech corpus and provide robust style/emotion/speaker representations for TTS.  
Furthermore, we integrate the learned representations into an improved VITS~\cite{Kim2021ConditionalVA} model and conduct experiments on a multi-domain dataset. Experimental results show that our proposed framework can synthesize diverse stylistic and emotional speech for a target speaker who does not have the target style or emotion in the training data. We suggest readers listen to our online demos~\footnote{Demo:\href{https://zxf-icpc.github.io/MSES/}{https://zxf-icpc.github.io/MSES/}}.

\section{PROPOSED APPROACH}
The proposed approach comprises an SRL module and a VITS model. Based on contrastive learning, the SRL module aims to extract disentangled style, emotion, and speaker representation from speech. The VITS model synthesizes speech conditioned on the extracted representation.

\begin{figure}[htb]

\begin{minipage}[b]{1.0\linewidth}
  \centering
  \centerline{\includegraphics[width=\textwidth]{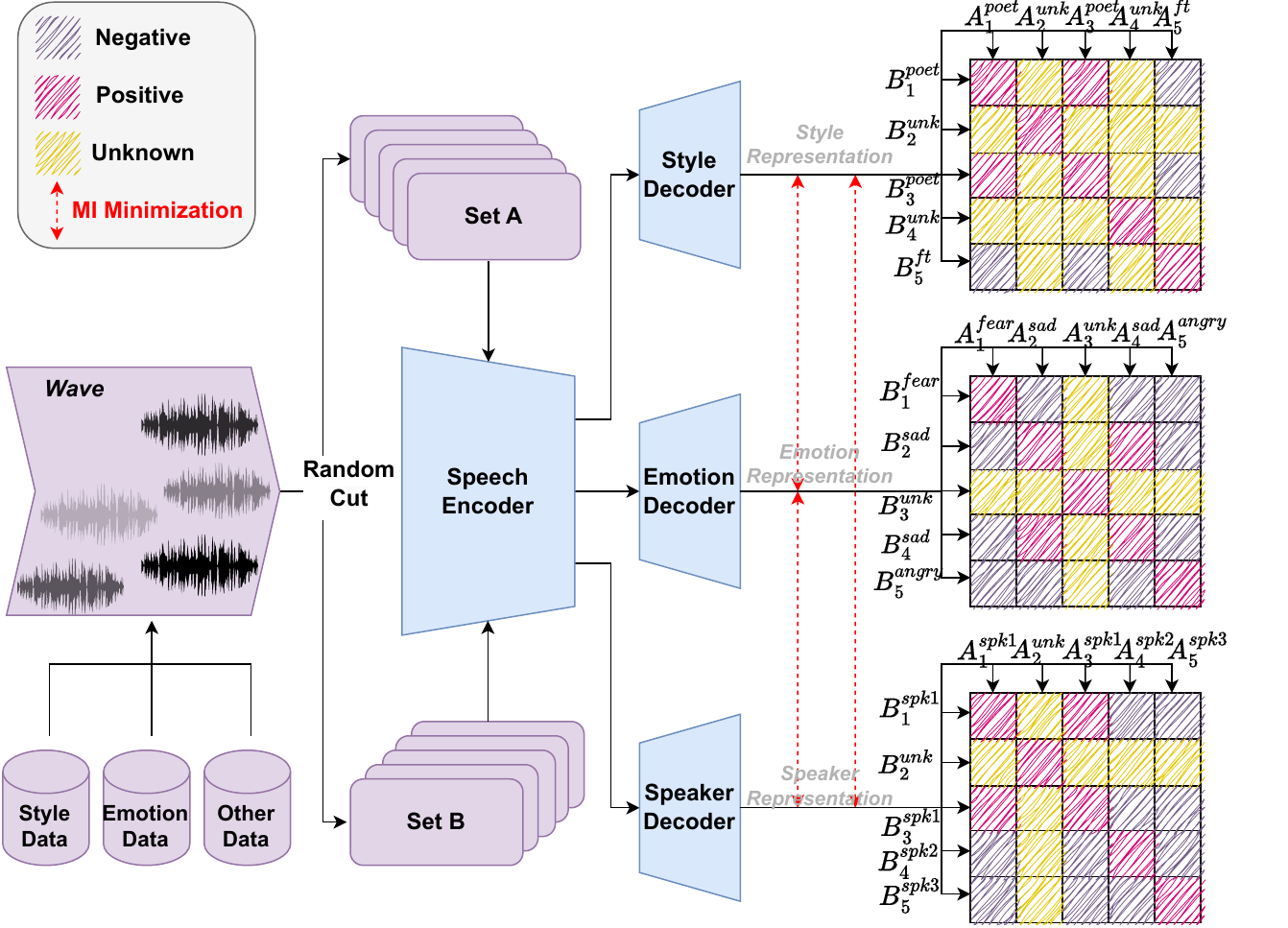}}
\end{minipage}
\caption{The architecture of speech representation learning module.}
\label{fig_1}
\end{figure}

\subsection{Semi-supervised positive and negative sample construction}

The key to contrastive learning lies in constructing positive and negative samples. Typically, Positive samples are constructed through data augmentation or data from the same category, while negative samples consist of data from different categories~\cite{surveyoncl, SIMCLR}. In our approach, the objective is to learn style, emotion, and speaker representations simultaneously. Therefore, multi-domain training data including labeled and unlabeled speech data is available to obtain these disentangled representations. With the diverse training data, we randomly split speech into speech slices and propose a positive and negative sample construction strategy at different levels. 
\begin{itemize}
    \item \textbf{Utterance level}: The style, emotion, and speaker timbre are regarded as global attributes, which change slowly alongside the time axis within each single utterance, so different slices of the same utterance can be treated as positive examples of each other.
    \item \textbf{Category level}: Speech samples from the same category exhibit highly correlated style, emotional expression, or speaker timbre. Therefore, speech slices of the same category are positive examples of each other, while speech slices of different categories are negative examples pairs. 
\end{itemize}

With the above utterance-level and category-level sample construction method, we further introduce a semi-supervised training strategy to the contrastive learning, which enables the SRL module to leverage abundant speech data and improve the robustness of the SRL module.
Specifically, unlabeled speech data can be used to construct positive samples at the utterance level, while the category-level sample pairs of unlabeled data are not involved during training due to the undefined relationships between the sample pairs. Moreover, randomly Selecting negative samples for unlabeled data will cause adverse effects~\cite{DBLP:journals/jstsp/ChoVMD22}. Labeled speech data can be leveraged to construct sample pairs at both utterance level and category level.

\subsection{Contrastive learning module}

As shown in Figure \ref{fig_1}, the SRL module consists of a speech encoder, a style decoder, an emotion decoder, and a speaker decoder. The speech encoder comprises a Hubert model and transformer blocks. The Hubert model is to extract features from speech for its exceptional performance across diverse downstream tasks~\cite{DBLP:journals/taslp/HsuBTLSM21}. Transformer blocks encode the Hubert features into three hidden features. These hidden features are then fed to the style, emotion, and speaker decoders to produce global style, emotion, and speaker representations, respectively. We normalize all style, emotion, and speaker representations to a hypersphere by \textit{l2-normalizing}. This normalization eliminates information related to the magnitude and reserves information related to the angular~\cite{DBLP:conf/smc/KimLLJL21}, effectively improving the supervision of cosine similarity.

During training, given $K$ speech waveforms, we randomly cut two speech slices from each speech waveform, forming two sets of speech slices, Set A and Set B. We calculate a $K \times K$ cosine similarity matrix $\hat{M}$ between the representation of Set A and Set B, where the value at the position of the $i^{th}$ row and $j^{th}$ column indicates the cosine similarity between the representation of the $i^{th}$ speech slice in Set A and $j^{th}$ speech slice in Set B. The ground truth matrix $M$ consists of -1, 0, and 1 values, where 1 represents positive, 0 represents negative, and -1 represents unknown. The SRL module generates $\hat{M}_{style}$, $\hat{M}_{emotion}$, and $\hat{M}_{speaker}$, and calculate the loss with corresponding the ground truth matrix $M_{style}$, $M_{emotion}$, and $M_{speaker}$, respectively.

We calculate the contrastive learning loss $\mathcal{L}_{con}$ between $\hat{y}$ in the cosine similarity matrix $\hat{M}$ and $y$ ground-truth matrix $M$. $\mathcal{L}_{con}$ takes Cross-entropy as the loss function as follows:
\begin{equation}
L_{con}(y, \hat{y}) = \begin{cases}
  -\log(\hat{y}) & \text{if } y = \text{positive}, \\
  -\log(1 - \hat{y}) & \text{if } y = \text{negative}, \\
  0 & \text{if } y = \text{unknown}.
\end{cases}
\end{equation}
Moreover, we further disentangle style, emotion, and speaker representation through mutual information (MI) minimization.
Given the random variables \textbf{u} and \textbf{v}, the MI is Kullback-Leibler (KL) divergence between their joint and marginal distributions
as $\text{I}(u,v) = D_{KL}(P(u,v); P(u)P(v))$. We adopt vCLUB~\cite{Cheng2020CLUBAC} to compute the upper bound of MI and calculate the MI loss $\mathcal{L}_{MI}$ as: 
{
\begin{equation} \label{eq1}
\begin{split}
 \mathcal{L}_{\mathrm{MI}}  = & {{\text{I}}(style, emotion)} + {{\text{I}}(emotion, speaker)} \\ & + {{\text{I}}(speaker, style)}
\end{split}
\end{equation}
}
Therefore, the final training objective of the SRL module $\mathcal{L}_{\mathrm{srl}}$ is as follows:
\begin{equation}
    \mathcal{L}_{\mathrm{srl}}= \mathcal{L}_{con} + \mathcal{L}_{\mathrm{MI}}
\label{eq:eq1}
\end{equation}

\subsection{Expressive VITS}

\begin{figure}[htb]

\begin{minipage}[b]{1.0\linewidth}
  \centering
  \centerline{\includegraphics[width=\textwidth]{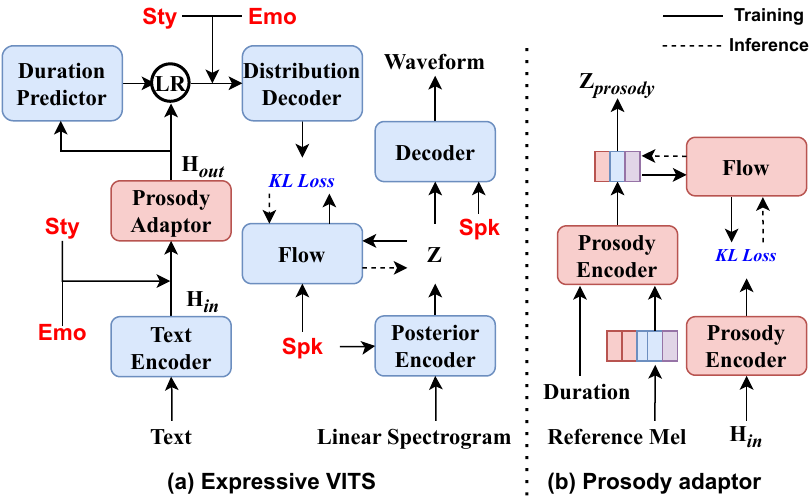}}
\end{minipage}

\caption{The architecture of multi-speaker expressive VITS.}
\label{fig_2}
\end{figure}

After training the SRL module, a VITS model is trained on conditioning of the extracted style, emotion, and speaker representations by contrastive learning. As shown in Figure \ref{fig_2}(a), we use VITS-CLONE~\cite{DBLP:journals/corr/abs-2207-06088} as the backbone for its excellent performance on expressive speech synthesis. Specifically, to improve the control ability, we replace the stochastic duration predictor and Monotonic Alignment Search (MAS) module in VITS with the duration predictor and length regulator in FastSpeech2~\cite{Ren2021FastSpeech2F}. Moreover, we use a flow-based prosody adaptor, as shown in Figure \ref{fig_2}(b), to capture fine-grained prosody variation of speech and improve the expressiveness of synthetic speech. The prosody adaptor encodes phoneme level $Z_{prosody}$ from the reference mel-spectrogram, which is added with the text encoder output $H_{in}$ to obtain $H_{out}$.

As shown in Figure \ref{fig_2}(a), we add the text encoder output and distribution decoder input with style and emotion representations to control style and emotional expression. Speaker representations are conditioned to the flow, posterior encoder, and decoder. The training objective is the same as that of CLONE. In inference, the SRL module extracts style, emotion, and speaker representation from reference speech, which are then sent to the VITS model with text to synthesize target speech.

\section{EXPERIMENTAL SETUPS}

\subsection{Datasets}

There are five corpora involved in the experiments. 1) \textbf{CN30S3} contains 18.5 hours of Chinese speech from 30 speakers, where each speaker has one to three styles, including poetry recitation, fairy tales, and storytelling - novels. 2) \textbf{CN3E6} contains 21.1 hours of Chinese speech from 30 speakers, and each speaker in this dataset has six emotions: anger, fear, happiness, sadness, surprise, and neutral. 3) \textbf{CN5U} has 5.8 hours of Mandarin speech from 5 speakers. 4) \textbf{EN5U} 31.3 hours of English speech from 5 speakers. 5) \textbf{MIXU} has 900 hours of Chinese and English speech collected from internal resources without annotations and transcripts.

For all recordings, we down-sample them into 24k Hz and set the frame and hop size to 1200 and 300, respectively. We cut recordings into 3 seconds of speech slices to construct the sample pairs of contrastive learning. For that speech of less than 3 seconds, we repeat them in the time axis until they are more than 3 seconds. We use a Bilingual TTS front end to encode Chinese text input into phonemes, tones, word boundaries, and prosodic boundaries while decoding English text input into phonemes. We leverage Pinyin and CMU-Dict as the phoneme set. The phoneme duration is obtained through an HMM-based force alignment model~\cite{Sjlander2003AnHS}. 

\subsection{Model configuration}

To validate the performance of our proposed approach, we implement the following systems:
\begin{itemize}
    \item \textbf{TSEW}~\cite{10095776}: A two-stage framework with a \textbf{T}ext-to-\textbf{S}tyle-and-\textbf{E}motion module and a style-and-emotion-to-\textbf{W}ave module. The former module predicts neural bottleneck features with style and emotion, while the latter predicts waveforms from bottleneck features conditioned on the speaker and emotion embedding.
    \item \textbf{SCVITS}~\cite{scser}: VITS model with a \textbf{S}upervised-\textbf{C}ontrastive learning module. Specifically, we replace Wav2Vec2.0 with Hubert for fair comparison and train three models to learn emotion, style, and speaker representations on the corresponding labeled dataset, respectively. We train the VITS model on these learned representations with the same configuration as the proposed approach.
    \item \textbf{Proposed}: The proposed framework with an SRL module and VITS model.

\end{itemize}

In our implementation, we use the Chinese-Hubert-Large~\footnote{HuBERT:\href{https://github.com/TencentGameMate/chinese_speech_pretrain/}{https://github.com/TencentGameMate/chinese\_speech\_pretrain}} to extract features from layer 6 to layer 18. We find that the Hubert features from a single layer, such as layer 6, perform worse than those from multiple layers in our experiments. Transformer blocks consist of 3 layers, 2 attention heads, an embedding dimension of 256, a feed-forward layer dimension of 1024, and a dropout of 0.2. The structure of style, emotion, and speaker decoders is the same and follows the structure of the reference encoder in the prior work~\cite{10095776}. The mutual information estimator also follows the settings of~\cite{10095776}. The backbone of expressive VITS keeps the same configuration as CLONE.

We train the SRL module with a batch size of 96 and the expressive VITS model with a batch size of 48. To balance the labeled and unlabeled data during the training of the SRL module, the batch is divided into four equal parts. One-fourth of the batch comes from style-labeled data, another-fourth comes from emotion-labeled data, and the third and fourth parts come from speaker-labeled and unlabeled data, respectively.

\begin{table*}[]
\centering
\caption{Results of monolingual subjective evaluation with 95$\%$ confidence interval and objective evaluation.}
\label{table1}
\begin{tabular}{l|cccc|cc}
\toprule
Model    & Naturalness↑ & Emotion Similarity↑ & Speaker Similarity↑ & Style Similarity↑  & CER(\%)↓  & SCS↑  \\ \midrule
TSEW & 3.94 $\pm$ 0.10            & 3.91 $\pm$ 0.11                   & 3.88 $\pm$ 0.07 & 3.85 $\pm$ 0.11    & 6.2          & 0.866               \\
SCVITS      & 3.97 $\pm$ 0.08            & 3.84 $\pm$ 0.10                   & 3.75 $\pm$ 0.11   & 3.69 $\pm$ 0.11    & 4.8            & 0.884             \\ 
Proposed & \textbf{4.09 $\pm$ 0.08}            & \textbf{3.96 $\pm$ 0.11}                   & \textbf{3.97 $\pm$ 0.07} & \textbf{4.03 $\pm$ 0.11}  & \textbf{3.0}           & \textbf{0.909} \\
\quad - MI      & 4.01 $\pm$ 0.09            & 3.88 $\pm$ 0.10                   & 3.90 $\pm$ 0.09 & 3.88 $\pm$ 0.11   & 4.5         & 0.893   \\ \bottomrule
\end{tabular}
\vspace{-10pt}
\end{table*}

\subsection{Evaluation metrics}
For monolingual TTS evaluation, given 20 reserved transcripts for each style, we generate samples respectively for each emotion category, resulting in 360 listening samples per person (20 texts $\times$ 3 styles $\times$ 6 emotions). We randomly select two speakers from CN5U as target speakers. For multilingual TTS evaluation, we add 20 English transcripts and generate samples for each emotion category, resulting in 120 listening samples. We randomly select a speaker from each of CN5U and EN5U as target speakers.

For subjective evaluation, We conduct Mean Opinion Score (MOS) experiments to evaluate speech naturalness and Similarity Mean Opinion Scores (SMOS) to evaluate emotion similarity, speaker similarity, and style similarity, respectively. Twenty volunteers with basic bilingual skills take part in the assessment. During the evaluation, participants are told to focus on specific aspects while ignoring others. For objective evaluation, we use an ECAPA-TDNN~\cite{Desplanques2020ECAPATDNNEC} model trained on 3,300 hours of Mandarin speech and 2,700 hours of English speech from 18,083 speakers to measure speaker cosine similarity (SCS). Moreover, we use an open-source U2++ conformer model provided by the WeNet community~\cite{Yao2021WeNetPO} to evaluate character error rate (CER), and word error rate (WER). The U2++ conformer model is trained on 10,000 hours of open-source Gigaspeech English data and WeNet Mandarin data, respectively.

\vspace{-10pt}
\section{EXPERIMENTAL RESULTS}
We first evaluate the performance of the proposed approach on monolingual corpora (CN30S3, CN3E6, and CN5U). Then, we conduct experiments on multilingual corpora (CN30S3, CN3E6, CN5U and EN5U), examining the effectiveness of the proposed approach in the cross-lingual transfer setting.

\subsection{Monolingual subjective evaluation}
As shown in Table~\ref{table1}, the proposed approach outperforms the compared models in terms of naturalness, emotion similarity, speaker similarity, and style similarity. TSEW gets the lowest naturalness, which we speculate that the error accumulation of BN prediction leads to this phenomenon. SCIVTS achieve higher naturalness but the lowest emotion, speaker, and style similarity. The style, emotion and speaker representations are extracted from three independent modules, which are probably entangled and lead to low similarity. These results show that our proposed approach obtains well-disentangled emotion, style, and speaker representations. Besides, the end-to-end VITS model avoids the error accumulation of intermediate representations, enabling flexible and natural expressive speech synthesis.

Moreover, to verify the advantages of the proposed method, we remove the mutual information estimator in the SRL module. As shown in Table~\ref{table1}, the model (-MI) still outperforms the compared models, showing the effectiveness of contrastive learning. With the mutual information estimator, the SRL module can better disentangle the style, emotion, and style representation, and the whole model obtains better performance.

\subsection{Monolingual objective evaluation}

As shown in Table~\ref{table1}, the proposed approach achieves the lowest CER, showing the robustness of the proposed framework. Moreover, the proposed approach obtains the highest SCS, indicating the speaker characteristics are well captured and disentangled. TSEW gets the worst CER, which we conjecture is due to the two-stage framework and the error accumulation of BN prediction. Moreover, SCVITS gets lower speaker cosine similarity, proving that the representation learned by SCVITS is not well-disentangled. Removing MI from the proposed framework leads to a performance decline in all objective evaluations, demonstrating the effectiveness of MI in achieving accurate pronunciation and high speaker cosine similarity. 

\begin{figure}[htb]
\centering

\begin{minipage}[b]{0.49\linewidth}
  \centering
  \includegraphics[width=\textwidth]{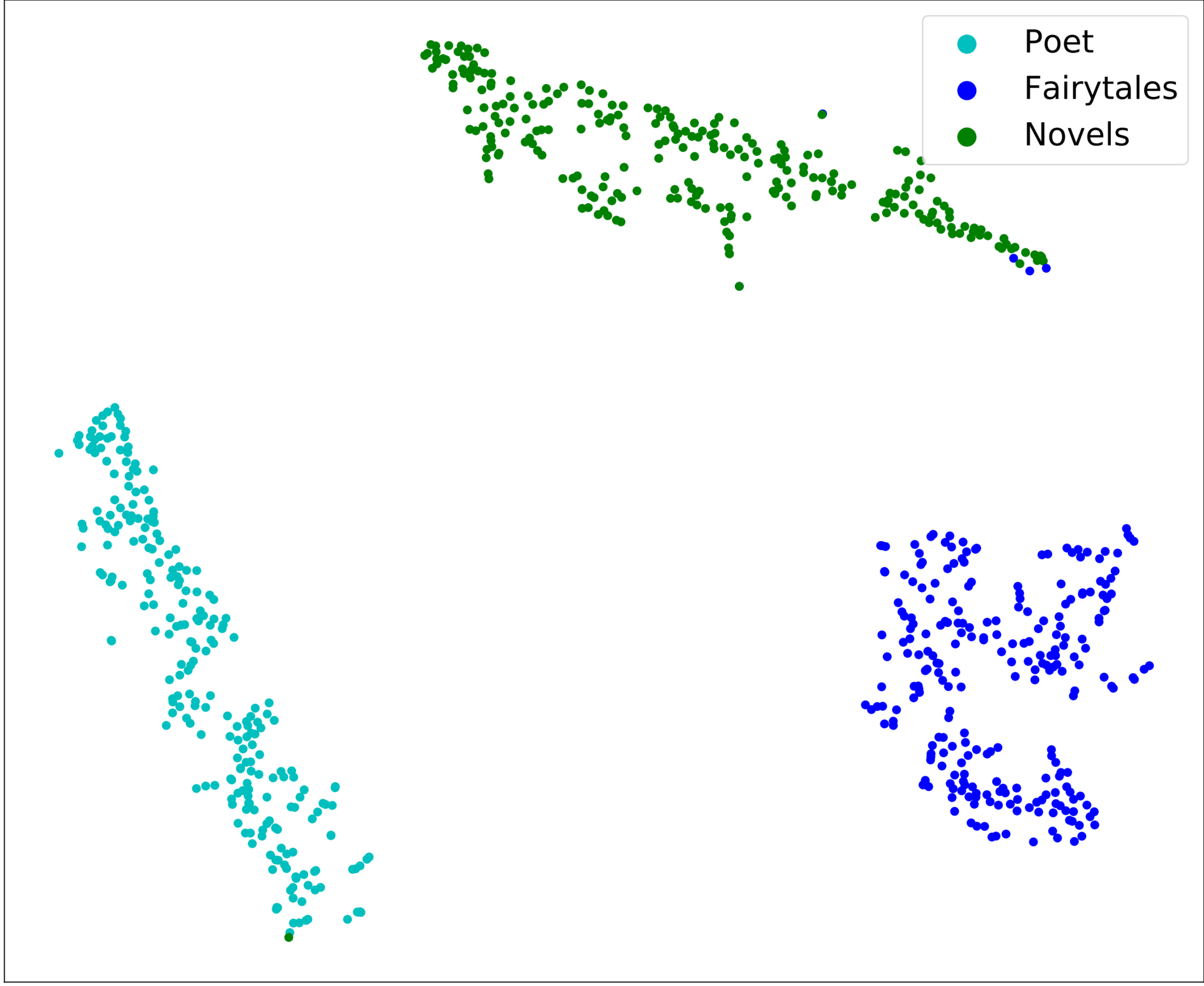}
  \label{fig:image1}
\end{minipage}
\hfill
\begin{minipage}[b]{0.49\linewidth}
  \centering
  \includegraphics[width=\textwidth]{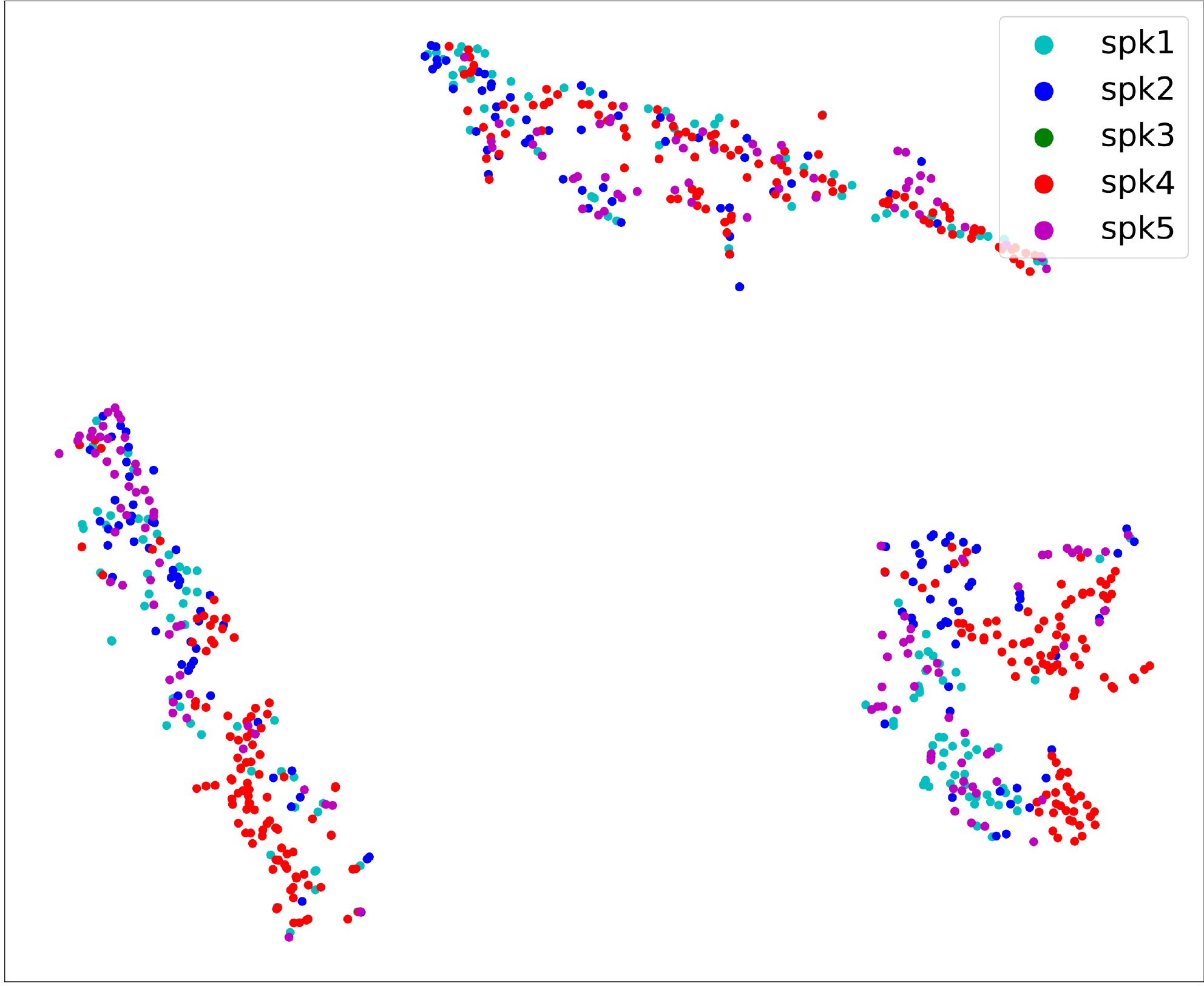}
  \label{fig:image2}
\end{minipage}
\begin{minipage}[b]{0.49\linewidth}
  \centering
  \includegraphics[width=\textwidth]{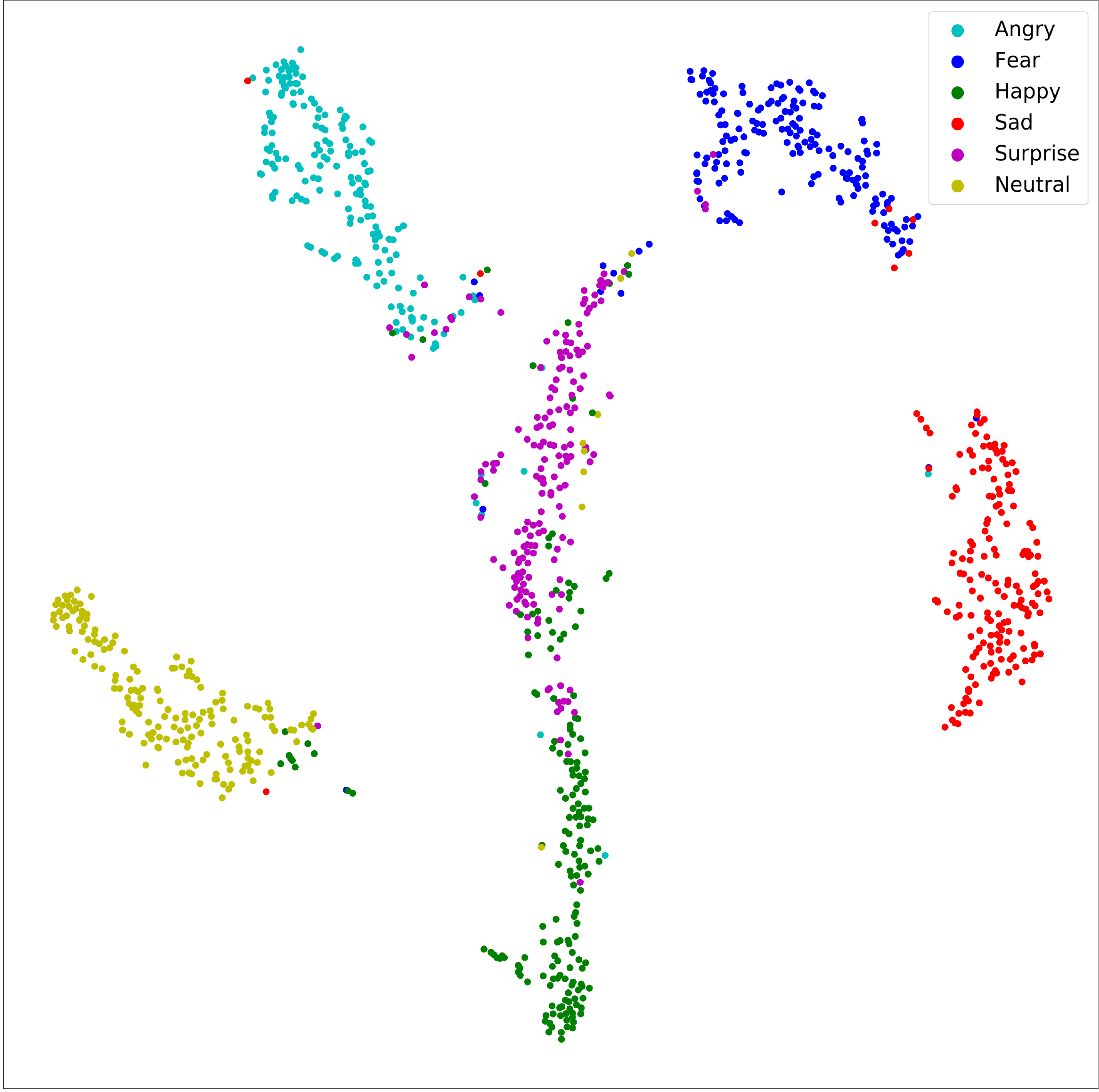}
  \label{fig:image3}
\end{minipage}
\hfill
\begin{minipage}[b]{0.49\linewidth}
  \centering
  \includegraphics[width=\textwidth]{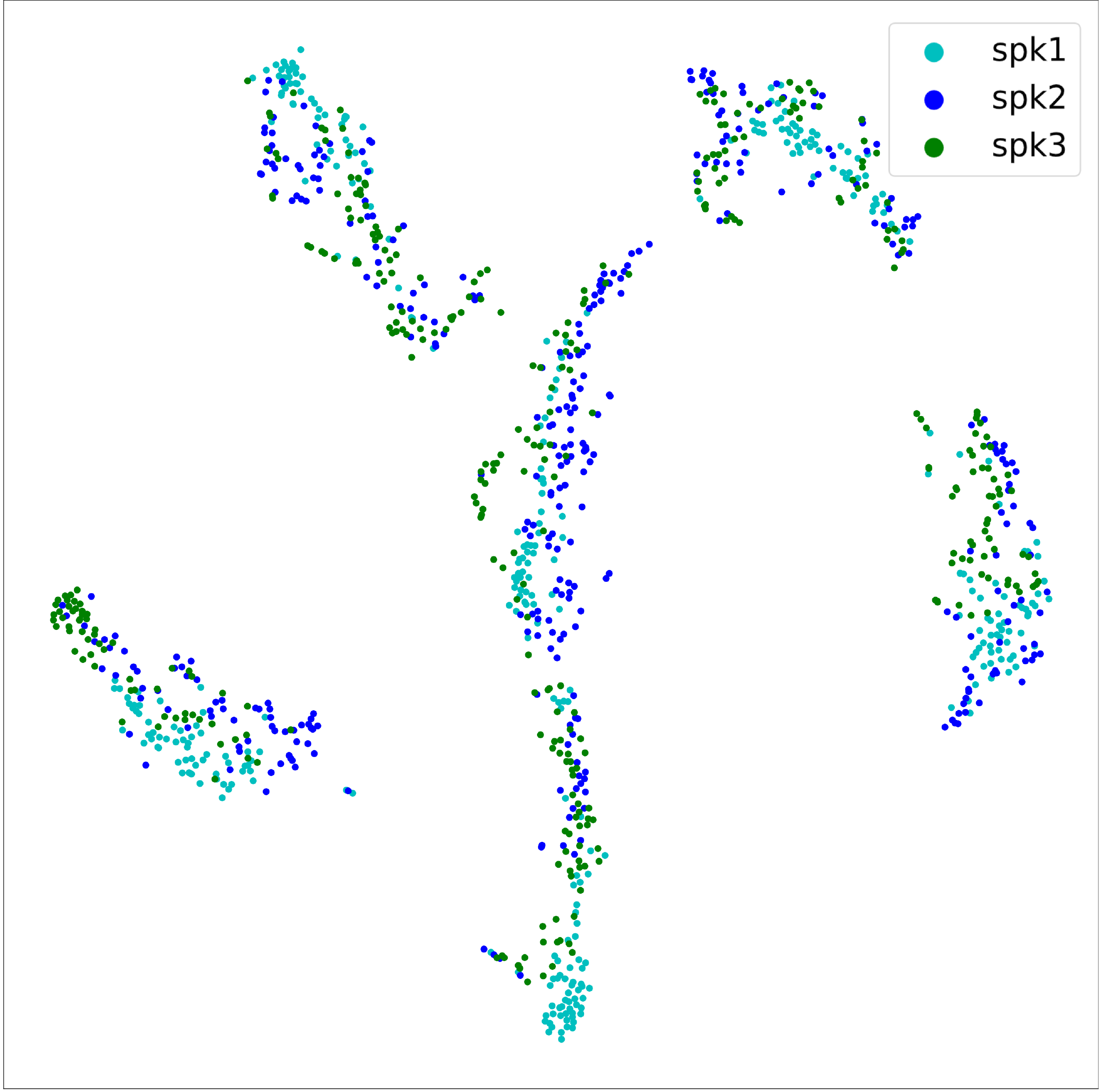}
  \label{fig:image4}
\end{minipage}

\caption{T-SNE visualization of style representation (above) and emotion representation (below). We color the results with the corresponding category (left) and speaker category (right).}
\label{fig:overall}
\end{figure}

Furthermore, to verify the effectiveness of the SRL module, we visualize the emotion and style representations through t-SNE~\cite{Maaten2008VisualizingDU}. One hundred fifty utterances reserved per emotion and 250 per style are adopted for test. As shown in Figure \ref{fig:overall}, the style and emotional representations are well clustered by corresponding categories, indicating the effectiveness of the SRL model. Moreover, the style and emotional representations can not be clustered by speaker identities, showing the good disentanglement between speaker and style, speaker and emotion.

\subsection{Multilingual subjective evaluation}

\begin{table*}[]
\centering
\caption{Results of multilingual subjective evaluation with 95$\%$ confidence interval and objective evaluation.}
\label{table2}
\begin{tabular}{l|cccc|ccc}
\toprule
Model    & Naturalness↑ & Emotion Similarity↑ & Speaker Similarity↑ & Style Similarity↑  & CER(\%)↓ & WER(\%)↓ & SCS↑  \\ \midrule
TSEW & 3.76 $\pm$ 0.12            & 3.87 $\pm$ 0.10                   & 3.61 $\pm$ 0.11 & 3.77 $\pm$ 0.10    & 6.8         &  9.7        & 0.847               \\
SCVITS      & 3.84 $\pm$ 0.08            & 3.80 $\pm$ 0.12                   & 3.70 $\pm$ 0.11   & 3.65 $\pm$ 0.12    & 5.7         &  4.9        & 0.838             \\ 
Proposed & 4.01 $\pm$ 0.07            & 3.92 $\pm$ 0.10                   & 3.90 $\pm$ 0.08 & 4.01 $\pm$ 0.09  & \textbf{3.9}         &  \textbf{2.7}        & 0.896 \\
\quad - MI      & 3.92 $\pm$ 0.10            & 3.84 $\pm$ 0.12                   & 3.77 $\pm$ 0.09 & 3.84 $\pm$ 0.12   & 5.2         &  3.6        & 0.852   \\        
\quad + MIXU  & \textbf{4.03 $\pm$ 0.08}            & \textbf{3.98 $\pm$ 0.11}                   & \textbf{3.95 $\pm$ 0.08} & \textbf{4.04 $\pm$ 0.10}  & 3.9         &  2.8        & \textbf{0.903}  \\ \bottomrule
\end{tabular}
\vspace{-10pt}
\end{table*}

The subjective evaluation results of multilingual TTS are shown in Table~\ref{table2}; all models exhibit a performance degradation compared to Table~\ref{table1}, revealing the challenges of cross-lingual expressive speech synthesis. However, the proposed approach demonstrates relatively minor degradation in performance during cross-lingual expressive speech synthesis, suggesting its capability to generate fluent and expressive foreign speech for a given target speaker. 
TSEW gets the lowest naturalness due to the unnatural pronunciation in synthetic English speech, which also affects the listeners’ judgment in speaker similarity evaluation. Primarily, BN in TSEW is extracted through a robust TDNN-F model trained with 30k hours of Chinese speech data, resulting in inaccurate pronunciation in English speech synthesized by TSWE. 
Moreover, SCVITS obtains a serious performance degradation. The supervised contrastive learning module of SCVITS can only be trained on labeled data, which means English speech is unseen during training and causes performance degradation.  
These results show the effectiveness of multi-level contrastive learning and semi-supervised training strategy in cross-lingual expressive speech synthesis.

Removing the mutual information estimator in the SRL module encounters a slight performance decline, which is consistent with monolingual evaluation. Besides, to evaluate the crucial role of the semi-supervised training strategy, we add the corpus MIXU during the SRL module training. With enlarged training corpora, the overall performance of the proposed system is improved, which means the wealth of variation in abundant unlabeled speech data helps capture more precise style, emotion and speaker characteristics.

\subsection{Multilingual objective evaluation}

As shown in Table~\ref{table2}, the proposed approach achieves the lowest CER and WER and the highest SCS, indicating the robustness of the proposed approach in cross-lingual expressive speech synthesis. TSEW gets the worst WER as BN is extracted through the TDNN-F model trained on Chinese speech data and the pronunciation is inaccurate.
Additionally, SCVITS fails to effectively address the challenge of emotion, style, and speaker entanglement in multilingual settings, yielding low SCS and high CER and WER.
Removing MI from the proposed framework leads to a performance decline in all objective evaluations while adding the corpus MIXU improves overall performance. 
These results confirm the observations from the subjective evaluation.

\begin{figure}[htb]
\centering

\begin{minipage}[b]{0.49\linewidth}
  \centering
  \includegraphics[width=\textwidth]{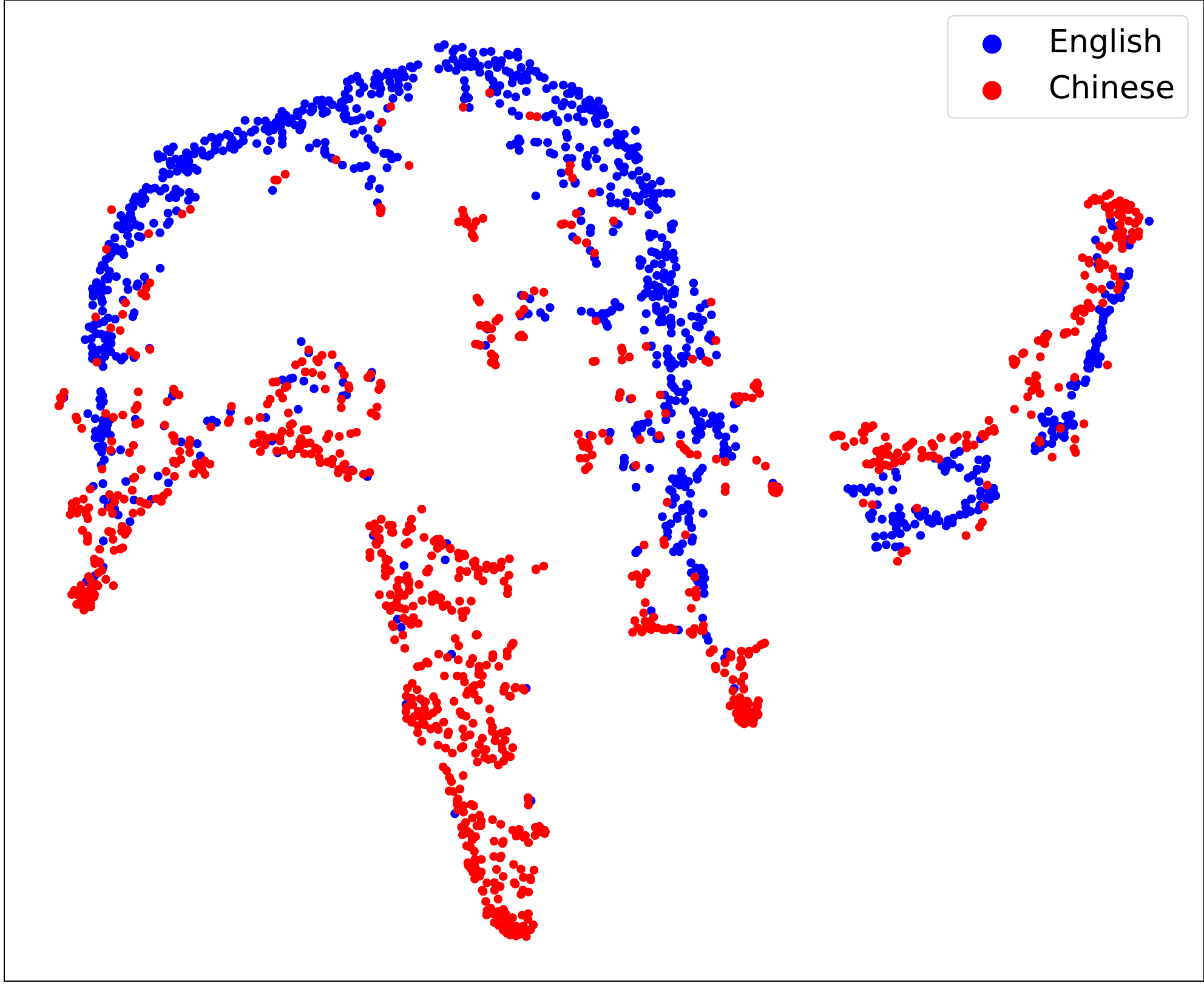}
  \label{fig:image1}
\end{minipage}
\hfill
\begin{minipage}[b]{0.49\linewidth}
  \centering
  \includegraphics[width=\textwidth]{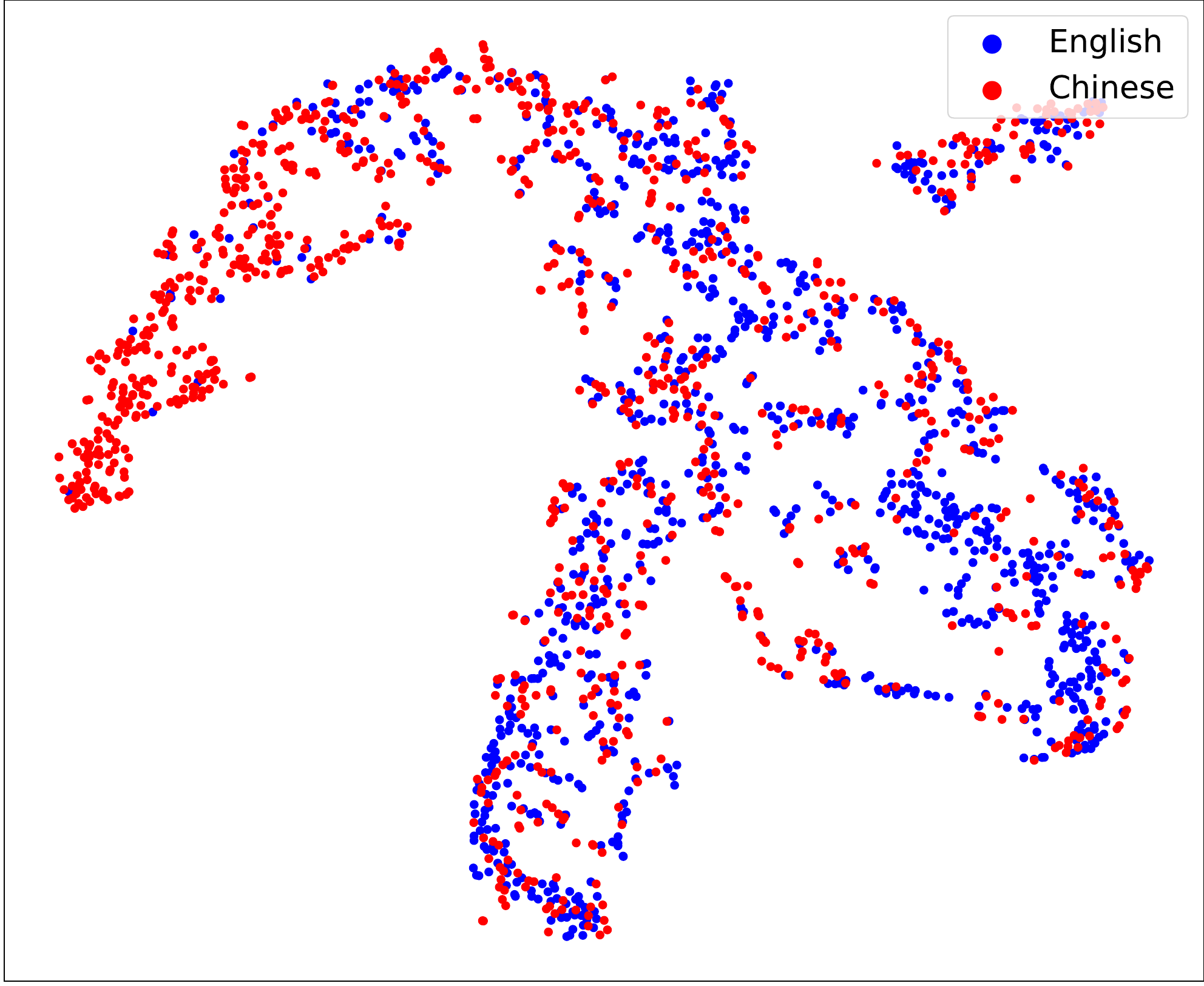}
  \label{fig:image2}
\end{minipage}

\caption{T-SNE visualization of style representation (left) and emotion representation (right) in multilingual settings.}
\label{fig2}
\end{figure}

Moreover, to study the relationship between style, speaker and language, we visualize multilingual emotion and style representations through t-SNE~\cite{Maaten2008VisualizingDU}. One thousand utterances reserved per language are adopted for test. As shown in Figure \ref{fig2}, the style representation tends to be language-specific while the emotion representation seems to be language-agnostic. We speculate that basic emotional expressions such as happiness and sadness are available in all languages~\cite{DBLP:conf/chi/DaiFM09, DBLP:conf/interspeech/SchullerMLR05}, causing the emotion representation to be language-agnostic. However, different manners of pronunciation in different languages lead to different speaking styles, which results in language-specific style representations~\cite{valle-x}.

\section{Conclusions}
\label{sec:majhead}
This paper aims to synthesize speech with the desired style and emotion for target speakers by transferring the style and emotion from reference speech recorded by other speakers.
We approach this challenging problem with a novel contrastive learning-based TTS framework. 
Specifically, this paper proposes a novel speech representation learning module based on contrastive learning, which constructs sample pairs at utterance and category levels and learns disentangled style, emotion, and speaker representations. Besides, we introduce a semi-supervised training strategy to the proposed framework, which leverages multi-domain data and helps learn robust representations.
We integrate the learned style, emotion, and speaker representation into an improved VITS model and conduct experiments on monolingual and multilingual datasets. Extensive experimental results demonstrate the proposed framework can synthesize speech with diverse speaking styles and emotions for a target speaker, even if the speaking style or emotion comes from another language.

\bibliographystyle{IEEEbib}
\bibliography{icme2023template}

\end{document}